\documentstyle[aps,prl,twocolumn,floats,epsfig]{revtex}

\begin{document}

\newfont{\sans}{cmss17 scaled 1000}
\newfont{\lsans}{cmss17 scaled 1200}
\newfont{\Lsans}{cmss17 scaled 1440}

\newcommand{\gsim}{
\,\raisebox{0.35ex}{$>$}
\hspace{-1.7ex}\raisebox{-0.65ex}{$\sim$}\,
}

\newcommand{\lsim}{
\,\raisebox{0.35ex}{$<$}
\hspace{-1.7ex}\raisebox{-0.65ex}{$\sim$}\,
}

\newcommand{\fractional}{ \mbox{frac} } 
\newcommand{\integer}{ \mbox{int} }

\bibliographystyle{prsty}

\title{
\begin{flushleft}
\parbox[t]{6cm}
{
{\rm Z. Phys. B 102, 283--287 (1997) }
}
\hfill
\parbox[t]{4.6cm}
{
\begin{flushleft}
\vspace{-5mm}
\_\hrulefill\\ 
{\sans ZEITSCHRIFT\\ F\"UR PHYSIK B}\\ 
%\copyright $\;$ {\rm Springer-Verlag 1997}\\
Author's version 1997\\
\vspace{-4mm}
\_\hrulefill\\
\end{flushleft} 
}\\
\vspace{2cm} 
{\LARGE\bf
Susceptibilities and correlation functions\\ 
of the anisotropic spherical model\\ 
}      
\bigskip
{\bf D.~A. Garanin}\\
\medskip
{\small
I. Institut f\"ur Theoretische Physik, Universit\"at Hamburg,
Jungiusstr. 9, D-20355 Hamburg, Germany\\
(e-mail: garanin@physnet.uni-hamburg.de)\\
}
\bigskip\bigskip
{\small Received: 28 August 1996}
\end{flushleft}
}

\maketitle

\bigskip

{\bf Abstract.} \hspace{1mm}
The static transverse and longitudinal correlation functions (CF) of a 
3-dimensional ferromagnet are calculated 
for the exactly solvable anisotropic spherical model (ASM) determined as 
the limit $D\to\infty$ of the classical $D$-component vector model.
The results are {\em nonequivalent} to those for 
the standard spherical model of Berlin and Kac even in the isotropic case.
Whereas the transverse CF has the usual Ornstein-Zernike form for 
small wave vectors, the longitudinal CF shows a nontrivial behavior 
in the ordered region caused by spin-wave fluctuations.
In particular, in the isotropic case below $T_c$ one has 
$S_{zz}({\bf k})\propto 1/k$ (the result of the spin-wave theory) for 
$k\lsim \kappa_m \propto T_c-T$.

\begin{flushleft}
\_\hrulefill\\ 
\end{flushleft}
The spherical model (SM) of Berlin and Kac \cite{berkac52} 
was believed many years to be the only model in the statistical theory
of magnetism, 
which is exactly solvable in 3 dimensions.
The other model, which is the limit $D\to\infty$ 
of the classical $D$-component vector model introduced by Stanley 
\cite{sta68prl}, 
has the same partition function in the homogeneous case \cite{sta68pr}, 
and for this reason the latter was not considered as an independent 
model by many researchers and received relatively little attention.

However, the $D=\infty$ model is nonequivalent to the standard 
spherical model of Berlin and Kac and goes beyond it in many respects.
First, it can be easily generalized for anisotropic systems \cite{oka70}.
Second, it does not use the global spin constraint, which leads to unphysical 
results in spatially inhomogeneous situations. 
In particular, in the SM the Curie temperature $T_c$ 
of a 4-dimensional ferromagnetic film 
with free boundary conditions, which is infinite in 3 dimensions 
and finite in the 4th dimension, has been found 
by Barber and Fisher \cite{barfis73} to be a non-monotonous function 
of the number of its layers $N$.
To the contrast, an improved version of the spherical model using 
separate spin constraints in each layer \cite{cosmazmih76} leads in the case 
mentioned above to the monotonically increasing $T_c(N)$, as it should be.
However, even this improved version of the SM 
fails on physically meaningful 
3-dimensional ferromagnetic films, since the latter are 
2-dimensional objects and without anisotropy their
$T_c$ is zero for any finite thickness.
An adequate description of spatially inhomogeneous and low-dimensional 
ferromagnetic systems on the ``spherical'' level can be archieved only with 
the use of the anisotropic $D=\infty$ model, which can be called 
also anisotropic spherical model (ASM).
This model was recently applied to study the dimensional crossover 
of $T_c(N)$
of anisotropic ferromagnetic films \cite{gar96jpal} and the effects of 
thermal fluctuations in the Bloch-wall phase transition \cite{gar96jpa}.

Here it will be shown that the ASM deviates from the standard SM
even in the spatially homogeneous case and even in the isotropic limit, if 
the spin-spin correlation functions are concerned.
Indeed, below $T_c$ in the $D=\infty$ model
there are two different --- longitudinal and 
transverse --- correlation functions, whereas there is only one CF in the SM.
A less trivial reason for the difference between two models relies on the 
fact that a 
wave-vector-dependent CF is proportional to the appropriate 
susceptibility, which is the linear response to a 
{\em spatially inhomogeneous} sinusoidal field.
In this case, as was argued above, the global spin constraint used by the SM 
modifies the results.
As we shall see, the ${\bf k}$-dependent longitudinal CF of the ASM has a 
nontrivial non-Ornstein-Zernike form below $T_c$, which is similar to that 
following from the lowest-order spin-wave theory for $T \ll T_c$
\cite{morkaw62,patpok73,lovtro91}.
For the uniaxially anisotropic ferromagnetic model the ASM yields finite 
longitudinal and transverse susceptibilities and correlation lengths in the 
ordered region.

The anisotropic generalization of the classical $D$-component vector model 
of Stanley \cite{sta68prl} can be described by the Hamiltonian
%
%\marginpar{dham}
%
\begin{equation}\label{dham}
{\cal H} = -{\bf H}\sum_{i}{\bf m}_i -
 \frac{1}{2}\sum_{ij}J_{ij}\sum_{\alpha=1}^D \eta_\alpha 
m_{\alpha i} m_{\alpha j} ,
\end{equation}
where $|{\bf m}_i|=1$ and $\eta_\alpha\leq 1$ are anisotropy coefficients.
Here we consider the uniaxial model with $\eta_1 \equiv \eta_z =1$ and 
$\eta_\alpha \equiv \eta \le 1$ for $\alpha\ge 2$.
The model (\ref{dham}) can be conveniently treated by the classical spin 
diagram technique \cite{garlut84d,gar94jsp,gar96prb}, which allows 
classification of diagrams in powers of $1/D$ for $D\gg 1$
\cite{gar94jsp}.
The equation of state $m(H,T)$ of the anisotropic spherical model, where
$m\equiv \langle m_z \rangle$, is contained in the diagram series 
corresponding to the self-consistent Gaussian approximation (SCGA)  
\cite{garlut84d,gar96prb}, which becomes exact in the limit $D\to\infty$.
In terms of the dimensionless variables $\theta \equiv T/T_c^{\rm MFA}$ and 
$h \equiv H/J_0$, where $T_c^{\rm MFA}= J_0/D$ is the mean-field transition 
temperature and $J_0$ is the zero Fourier
component of the exchange interaction, one comes to the system of equations 
for the magnetization $m$ and the gap parameter $G$
\cite{garlut84d,gar96prb}
%
%\marginpar{shhereq}
%
\begin{eqnarray}\label{sphereq}
&&
G = \frac{ m }{ m + h } , \nonumber\\ 
&&
\theta G P(\eta G) = 1 - m^2 . 
\end{eqnarray}
Here
%
%\marginpar{pgen}
%
\begin{equation}\label{pgen}
P(X) \equiv 
v_0\!\!\!\int\!\!\!\frac{d{\bf q}}{(2\pi)^3}
\frac{1}{1-X\lambda_{\bf q}} ,
\end{equation}
where
$v_0$ is the unit cell volume and $\lambda_{\bf q} \equiv J_{\bf q}/J_0$. 
In the long-wavelength limit $\lambda_{\bf q} \cong 1 - \alpha q^2$, 
where $\alpha \sim a_0^2$ and $a_0$ is the lattice spacing.
For 3-dimensional lattices with the nearest neighbour interactions 
the integral $P(X)$ has the following properties:
%
%\marginpar{plims}
%
\begin{equation}\label{plims}
P(X) \cong \left\{
\begin{array}{ll}
1 + X^2/z,                       & X   \ll 1               \\
W - c_0\,(1-X)^{1/2},            & 1-X \ll 1 ,  
\end{array} 
\right. 
\end{equation}
where $z$ is the number of nearest neighbors,
$W$ is the Watson integral, and $c_0=v_0/(4\pi\alpha^{3/2})$. 
For the simple cubic (sc) lattice 
$v_0 = a_0^3$ and $\alpha = a_0^2/6$, hence $c_0 = (2/\pi)(3/2)^{3/2}$.
The solution of the system of equations (\ref{sphereq}) simplifies for zero 
field, where below $T_c$ one has $G=1$ (the zero spin-wave gap) and
%
%\marginpar{mspher}
%
\begin{equation}\label{mspher}
m = (1-\theta/\theta_c)^{1/2}, \qquad 
\theta \le \theta_c = 1/P(\eta) .
\end{equation}
It can be seen that in the isotropic case, $\eta=1$, 
the value of the phase transition temperature  
$\theta_c$ reduces to the well-known result $\theta_c=1/W$ \cite{berkac52}.
Using (\ref{sphereq}) one can calculate the longitudinal susceptibility 
$\chi_z \equiv \partial m / \partial H$.
The zero-field reduced susceptibility, 
$\tilde\chi_z \equiv J_0\chi_z$, has the form
%
%\marginpar{chiz}
%
\begin{equation}\label{chiz}
\renewcommand{\arraystretch}{2}
\tilde\chi_z = \left\{
\begin{array}{ll}
\displaystyle
\frac{ G }{ 1 - G } ,                            & \theta > \theta_c   \\
\displaystyle
\frac{\theta}{2m^2} [\eta P'(\eta) + P(\eta)] ,  & \theta < \theta_c , 
\end{array} 
\right. 
\end{equation}
where $P'(X)\equiv dP(X)/dX$ and $G$ satisfies the equation 
$\theta G P(\eta G) = 1$ above $\theta_c$.
Solving this equation near $\theta_c$ in the linear approximation in 
$1-G\ll 1$, using (\ref{mspher}), and introducing the reduced temperature 
variable 
%
%\marginpar{epsdef}
%
\begin{equation}\label{epsdef}
\epsilon \equiv \frac{ \theta_c }{ \theta } - 1 ,
\end{equation}
one can rewrite (\ref{chiz}) in the form
%
%\marginpar{chiz1}
%
\begin{equation}\label{chiz1}
\tilde\chi_z \cong I(\eta) \left\{
\begin{array}{lll}
(-\epsilon)^{-1},  & \theta > \theta_c \\
(2\epsilon)^{-1},  & \theta < \theta_c , 
\end{array} 
\right. 
\end{equation}
where 
%
%\marginpar{ieta}
%
\begin{equation}\label{ieta}
I(\eta) \equiv 1 + \frac{\eta P'(\eta)}{P(\eta)}
\cong
\left\{
\begin{array}{ll}
\displaystyle
1 + 2\eta^2/z ,       & \eta \ll 1
\\
\displaystyle
\frac{c_0}{2P(\eta)}\frac{1}{\sqrt{1-\eta}} , 
& 1-\eta \ll 1 . 
\end{array}
\right.
\end{equation}
The first line of (\ref{chiz1}) is valid for the weakly-anisotropic model,
$1-\eta\ll 1$, in the narrow temperature interval
%
%\marginpar{eps*}
%
\begin{equation}\label{eps*}
-\epsilon \ll \epsilon^* \equiv \frac{c_0}{4W}\sqrt{1-\eta} \ll 1 ,
\end{equation}
whereas the second one is valid in the whole region below $\theta_c$.
The latter is the reason to define $\epsilon$ in the non-standard form 
(\ref{epsdef}).
It can be seen from (\ref{chiz1}) that the critical behaviour of the 
longitudinal susceptibility 
in the ASM is the same as that in the mean field approximation (MFA), 
including the famous ratio of critical amplitudes 2. 
This equivalence holds for all the critical indices \cite{garlut84d}, 
since for $\eta<1$ the
square-root singularity of $P(X)$ at $X=1$ is suppressed.
In the extreme case of the ``spherical Ising model'', $\eta=0$,
the mean field approximation for the ASM becomes exact, since the 
fluctuations of the transverse spin components die out with the transverse 
coupling $\eta$ and the influence of longitudinal fluctuations 
vanishes in the spherical limit, $D\to\infty$.
For the anisotropic model, $\eta<1$, the quantity $\tilde\chi_z$ is finite in 
the ordered region, vanishing at $\theta=0$ and diverging at 
$\theta = \theta_c$.
On the contrary, 
the longitudinal susceptibility of the isotropic model, 
$\eta=1$, diverges as $\tilde\chi_z\propto (-\epsilon)^{-2}$ above $\theta_c$ 
and is infinite in the whole region below $\theta_c$ due to the 
Goldstone-mode fluctuations.
In this case for $h\ll 1$ one has 
%
%\marginpar{msqrth}
%
\begin{equation}\label{msqrth}
\Delta m \equiv m(h) - m(0) \cong \frac{ \theta c_0 }{ m^{3/2} }h^{1/2},
\end{equation}
which coincides with the result of the standard spherical model and is in 
accord with the prediction of the lowest-order spin-wave theory 
\cite{holpri40}
for $\theta \ll \theta_c $.
A similar expression for a general $D$-component vector model, which 
contains the additional factor $1-1/D$, was 
derived by Fisher and Privman \cite{fispri85} from the scaling arguments.
Measuring the square-root singularity of magnetization (\ref{msqrth})
in systems with spontaneous breaking of a continuous symmetry is a 
difficult task, and it was done only resently on EuS 
\cite{koegoedompie94} and on EuO \cite{flogoekoe96}.

%\par
%
\begin{figure}[t]
\unitlength1cm
\begin{picture}(11,6)
\centerline{\epsfig{file=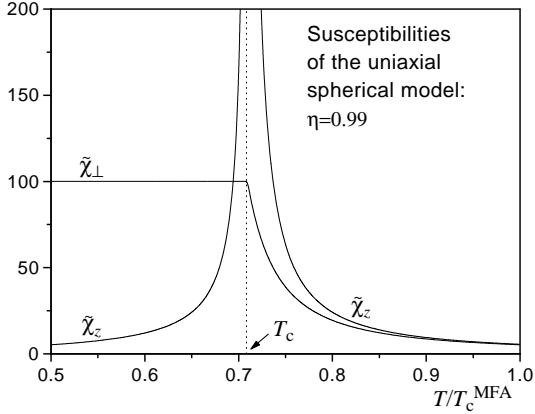,angle=-90,width=8cm}}
\end{picture}
%\par
%
\caption{ \label{scfchi}
Temperature dependences of the longitudinal and transverse susceptibilities 
of the uniaxial spherical model. 
}
\end{figure}

In the following we shall study the correlation functions
%
%\marginpar{cfdef}
%
\begin{equation}\label{cfdef}
S_{\alpha\alpha}({\bf r}) = 
\langle \Delta m_\alpha(0) \Delta m_\alpha({\bf r})\rangle,
\qquad \alpha = 1, 2, \ldots , D ,
%\qquad \Delta {\bf m}({\bf r}) \equiv {\bf m}({\bf r}) - 
%\langle m_z \rangle {\bf e}_z ,
%
\end{equation}
which are translationally invariant in the spatially homogeneous case
considered here.
The fluctuation-dis\-si\-pa\-ti\-on theorem relates wave-vector-dependent CFs 
to the appropriate susceptibilities as 
$\chi_\alpha({\bf k})=\beta S_{\alpha\alpha}({\bf k})$, 
where $\beta \equiv 1/T$.
Both longitudinal and transverse CFs enter the self-consistent Gaussian 
approximation and were (approximately) calculated in 
\cite{garlut84d,gar96prb}.
Since the SCGA becomes exact in the limit $D\to\infty$, the transverse CFs 
describing in the SCGA fluctuations of $D-1$ transverse spin components 
become exact, too, 
and with the use of the results of \cite{gar96prb} one obtains
%
%\marginpar{chitr}
%
\begin{equation}\label{chitr}
\tilde\chi_\perp({\bf k}) = \frac{D}{\theta} S_\perp({\bf k}) =
\frac{ G }{ 1 - \eta G \lambda_{\bf k} }
\end{equation}
in the whole temperature range [cf. (\ref{chiz})].
One can see that for $h=0$ below $\theta_c$, where $G=1$, the homogeneous
transverse susceptibility, 
$\tilde\chi_\perp \equiv \tilde\chi_\perp(0)$,
diverges in the isotropic limit, $\eta \to 1$, and remains a finite constant
for the uniaxially-anisotropic model.
Comparing (\ref{chiz1}) and (\ref{chitr}), 
one finds that both susceptibilities 
become equal to each other at the characteristic temperature slightly below 
$\theta_c$, which is determined by $\epsilon=\epsilon^*$ 
[see (\ref{epsdef}) and (\ref{eps*})].
The temperature dependences of the longitudinal and transverse 
susceptibilities for the weakly-anisotropic spherical model are represented 
in Fig. \ref{scfchi}.
One can see that even a small anisotropy, as $1-\eta = 10^{-2}$, has a 
profound influence on $\tilde\chi_z$ below $\theta_c$.
%\par
%
\begin{figure}[t]
\unitlength1cm
\begin{picture}(11,6)
\centerline{\epsfig{file=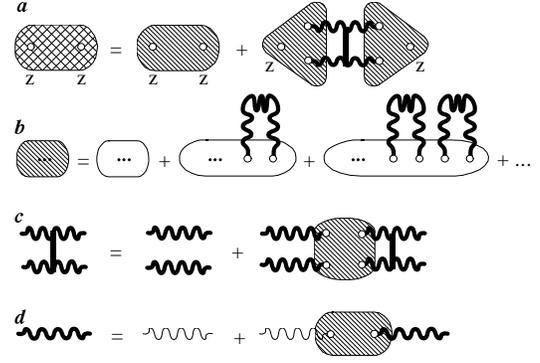,angle=0,width=12cm}}
\end{picture}
%\par
%
\caption{ \label{scfscf}
(a) The compact part $\hat\Lambda_{zz}({\bf k})$ of the longitudinal CF 
$S_{zz}({\bf k})$;
(b) block summation for one-site spin cumulants 
$\tilde\Lambda_{\ldots}$ 
renormalized by Gaussian fluctuations of the molecular field; 
(c) ladder equation for the transverse four-spin correlation line 
$\tilde V_{\bf k}$;
(d) Dyson equation for the renormalized transverse interaction 
$\eta\beta\tilde J_{\bf q}$.
Unlabeled small circles denote transverse spin components.
}
\end{figure}

Unlike the transverse CF given by (\ref{chitr}), the exact longitudinal CF, 
$S_{zz}({\bf k})$, cannot be determined from the SCGA in the spherical limit, 
since for $D\to\infty$ the fluctuations of {\em the single} longitudinal spin 
component become nonessential and $S_{zz}({\bf k})$ disappears from the SCGA 
equations.
Therefore, the wave-vector-dependent longitudinal CF in the anisotropic 
spherical model should be considered separately, which is the main purpose of 
this paper.
With the help of the classical spin diagram technique 
\cite{garlut84d,gar94jsp,gar96prb} $S_{zz}({\bf k})$ determined by
(\ref{cfdef}) can be represented as
%
%\marginpar{cfdys}
%
\begin{equation}\label{cfdys}
S_{zz}({\bf k}) = 
\frac{ \hat\Lambda_{zz}({\bf k}) }
{ 1 - \hat\Lambda_{zz}({\bf k}) \beta J_{\bf k} },
\end{equation}
where $\hat\Lambda_{zz}({\bf k})$ is the compact (irreducible) part of 
$S_{zz}({\bf k})$ given by the diagrams, which cannot be cut by the one 
longitudinal interaction line $\beta J_{\bf k}$.
The quantity $\hat\Lambda_{zz}({\bf k})$ is in turn given in the limit 
$D \to \infty$ by the set of diagrams represented in Fig. \ref{scfscf}.
Such a choice of diagrams is based on the arguments of \cite{gar94jsp}.
More technical details can be found in \cite{gar96jsp}, where the transverse 
CF, $S_{\alpha\alpha}({\bf k})$, $\alpha \geq 2$, 
was calculated up to the first order in $1/D$ 
for low-dimensional ferro- and antiferromagnets in magnetic field.
The analytical form of $\hat\Lambda_{zz}({\bf k})$ of Fig. \ref{scfscf}a
reads
%
%\marginpar{Lamzk}
%
\begin{equation}\label{Lamzk}
\hat\Lambda_{zz}({\bf k}) = 
\tilde\Lambda_{zz} + \tilde\Lambda_{\alpha\alpha z}^2 \tilde V_{\bf k}.
\end{equation}
Here in the limit $D\to\infty$ one has \cite{gar96jsp}
%
%\marginpar{Lamztil}
%
\begin{equation}\label{Lamztil}
\tilde\Lambda_{zz} = \tilde\Lambda_{\alpha\alpha} 
+ \tilde\Lambda_{\alpha\alpha\beta\beta} \xi^2,
\qquad 
\tilde\Lambda_{\alpha\alpha z} = 
\tilde\Lambda_{\alpha\alpha\beta\beta} \xi ,
\end{equation}
where $\alpha\neq\beta\neq z$, 
%
%\marginpar{Lamtil}
%
\begin{equation}\label{Lamtil}
\tilde\Lambda_{\alpha\alpha} = \frac{ \theta G }{ D },
\qquad
\tilde\Lambda_{\alpha\alpha\beta\beta} = 
- \left( \frac{ \theta G }{ D } \right)^3 \frac{ 1 }{ 1 - \theta G/2 },
\end{equation}
$G$ satisfies the system of equations (\ref{sphereq}), and
%
%\marginpar{xidef}
%
\begin{equation}\label{xidef}
\xi \equiv \beta (H + mJ_0) = \frac{ D }{ \theta }(h+m) 
= \frac{ Dm }{ \theta G }
\end{equation}
is the temperature-normalized molecular field.
The quantity $\tilde V_{\bf k}$ in (\ref{Lamzk}) is the solution of 
the ladder equation Fig. \ref{scfscf}c and has the form
%
%\marginpar{Vtil}
%
\begin{equation}\label{Vtil}
\tilde V_{\bf k} = 
\frac{ V_{\bf k} }{ 1 - \tilde\Lambda_{\alpha\alpha\beta\beta} V_{\bf k} },
\end{equation}
where $\tilde\Lambda_{\alpha\alpha\beta\beta}$ is given by (\ref{Lamtil}),
%
%\marginpar{Vdef}
%
\begin{equation}\label{Vdef}
V_{\bf k} = \frac{ D-1 }{ 2 }
\,v_0\!\!\!\int\!\!\!\frac{ d{\bf q} }{ (2\pi)^3 }\,
\eta\beta\tilde J_{\bf q} \, \eta\beta\tilde J_{\bf k-q},
\end{equation}
and the renormalized transverse interaction $\eta\beta\tilde J_{\bf q}$
determined by Fig. \ref{scfscf}d reads
%
%\marginpar{Jtil}
%
\begin{equation}\label{Jtil}
\eta\beta\tilde J_{\bf q} = \frac{ \eta\beta J_{\bf q} }
{ 1 - \tilde\Lambda_{\alpha\alpha} \eta\beta J_{\bf q} }.
\end{equation}
The factor $D-1$ in (\ref{Vdef}) results from the summation over 
transverse spin components in Fig. \ref{scfscf}a or in Eq. (\ref{Lamzk}), as 
well as in the ladder equation Fig. \ref{scfscf}c.
Such a factor does not appear, if one tries to take into account the similar 
diagrams for the {\em transverse} CF 
$S_\perp({\bf k})\equiv S_{\alpha\alpha}({\bf k})$, $\alpha \geq 2$, 
and thus these diagrams vanish 
in the spherical limit and $S_\perp({\bf k})$ has the
trivial Ornstein-Zernike form (\ref{chitr}).

\begin{figure}[t]
\unitlength1cm
\begin{picture}(11,6)
\centerline{\epsfig{file=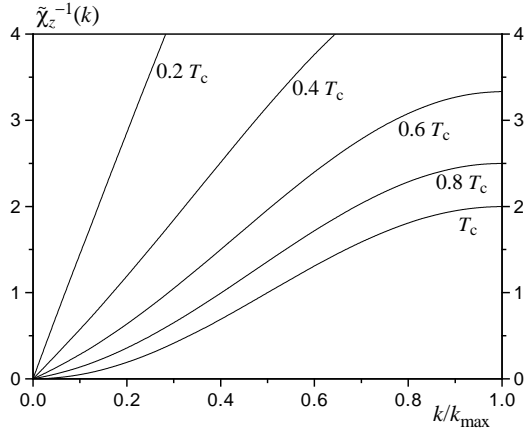,angle=-90,width=8cm}}
\end{picture}
%\par
%
\caption{ \label{scfkdep}
The wave vector dependences of the inverse longitudinal susceptibility 
$\tilde\chi_z^{-1}({\bf k})$ of the isotropic spherical ferromagnet 
along the [111] direction of ${\bf k}$
for different temperatures in the ordered region.
}
\end{figure}

Combining now Eqs. (\ref{cfdys})-(\ref{Jtil}), 
one comes after simplifications to the final result
%
%\marginpar{chizk}
%
\begin{equation}\label{chizk}
\tilde\chi_z^{-1}({\bf k}) = G^{-1} - \lambda_{\bf k} 
+ \frac{ 2m^2 }{ \theta G^2 } \frac{ 1 }{ r_{\bf k} } ,
\end{equation}
where
%
%\marginpar{rkdef}
%
\begin{equation}\label{rkdef}
r_{\bf k} = 
v_0\!\!\!\int\!\!\!\frac{ d{\bf q} }{ (2\pi)^3 }\,
\frac{ 1 }{ 1 - G\eta\lambda_{\bf q} }
\frac{ 1 }{ 1 - G\eta\lambda_{\bf k-q} } .
\end{equation}
The result similar to (\ref{chizk}), in which, however, only the last 
term is present, was obtained earlier 
\cite{morkaw62,patpok73,lovtro91}
within the lowest-order spin-wave theory well below $T_c$.
On the other hand, for the standard spherical model the correlation 
function has the trivial Ornstein-Zernike for all temperatures 
\cite{berkac52,joy72pt}.
The integral in (\ref{rkdef}) can be easily calculated for ${\bf k}=0$ and in 
the corner of the Brillouin zone, ${\bf k=b}$, for the sc lattice, where
$\lambda_{\bf b-q}=-\lambda_{\bf q}$.
The result has the form
%
%\marginpar{rk0b}
%
\begin{equation}\label{rk0b}
r_{\bf k} = P(\eta G) \left\{
\begin{array}{ll}
I(\eta G),                 & {\bf k}=0               \\
1,                         & {\bf k=b} ,  
\end{array} 
\right. 
\end{equation}
where $I(X)$ is given by (\ref{ieta}).
One can see that $\tilde\chi_z^{-1}(0)$ of (\ref{chizk}) is in accord
with the previously obtained expression (\ref{chiz})
for $h=0$ below $\theta_c$, where $G=1$. 
Under the same conditions $\tilde\chi_z^{-1}({\bf b})=\theta/(2\theta_c)$,
which can be used to control numerical calculations.
For the weakly anisotropic model in the case $1-\eta G \ll 1$ 
the integral (\ref{rkdef}) can be calculated analytically for small wave 
vectors, $a_0 k \ll 1$, which results in 
%
%\marginpar{rksm}
%
\begin{equation}\label{rksm}
r_{\bf k} \cong \frac{ c_0 }{ \sqrt{\alpha k^2} } 
\arctan \left( 
\frac{ 1 }{ 2 } \sqrt{ \frac{ \alpha k^2 }{ 1-\eta G } }
\right) .
\end{equation}
The last term in (\ref{chizk}) modifies the ${\bf k}$-dependence of 
$\chi_z({\bf k})$ below $\theta_c$ in the gapless case $\eta G=1$ in 
the range of small wave vectors, which shrinks if $\theta_c$ is approached 
from below.
Instead of the Ornstein-Zernike form $\chi_z({\bf k}) \propto 1/k^2$ one has
(cf. \cite{morkaw62,patpok73,lovtro91})
%
%\marginpar{kc}
%
\begin{equation}\label{kc}
\chi_z({\bf k}) \propto \frac{ 1 }{ k }, \qquad
k \ll \kappa_m = \frac{ 2m^2 }{ \theta } \frac{ 1 }{ c_0 \alpha^{1/2} }.
\end{equation}
The latter defines a length scale $\xi_m$, which exists only in the ordered 
region ($m>0$) and diverges at $\theta_c$ due to the vanishing of 
magnetization (\ref{mspher}):
%
%\marginpar{rc}
%
\begin{equation}\label{xim}
\xi_m \equiv \frac{ 1 }{ \kappa_m } = \frac{ \theta }{ 2m^2 } 
c_0\alpha^{1/2} .
\end{equation}
For the sc lattice $c_0\alpha^{1/2}=\frac{ 3 }{ 2\pi } a_0$ and 
hence $\xi_m=\frac{ 3\theta }{ 4\pi m^2 } a_0$.
The length $\xi_m$ is analogous to the ``bare'', i.e., the mean-field 
correlation length below $T_c$, which follows, in particular, from the 
Landau-Ginzburg phenomenological free energy.
This analogy is, however, not complete, since $\xi_m$ diverges at the 
actual transition temperature $T_c$ and not at $T_c^{\rm 
MFA}$ ($\theta=1$).
The crossover of $\chi_z({\bf k})$ at $k\sim\kappa_m$ for 
temperatures slightly below $T_c$ was described 
earlier with the help of the renormalization group approach \cite{ma76}.
The ${\bf k}$-dependences of $\tilde\chi_z^{-1}({\bf k})$ of 
(\ref{chizk}) in the isotropic case obtained with the help of numerical 
integration in (\ref{rkdef}) are represented in Fig. \ref{scfkdep}
for different temperatures below $T_c$. 
Curves of such a type for Heisenberg model systems such as
EuO and EuS could be observed, in principle, in
neutron scattering experiments, but such experiments were not carried 
out up to now. 
%\par
%

\begin{figure}[t]
\unitlength1cm
\begin{picture}(11,6)
\centerline{\epsfig{file=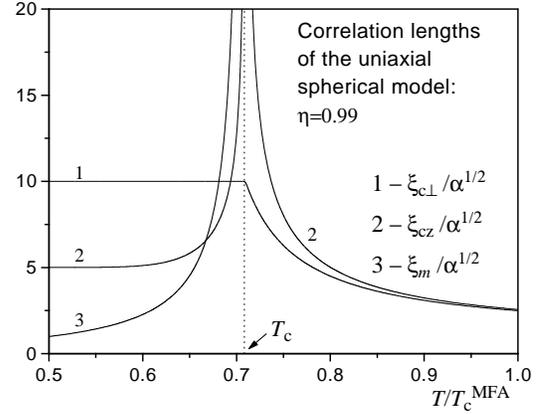,angle=-90,width=8cm}}
\end{picture}
%\par
%
\caption{ \label{scfxi}
Temperature dependences of the transverse, $\xi_{c\perp}$, and
longitudinal, $\xi_{cz}$,  correlation 
lengths, as well as the characteristic length $\xi_m$, 
in the uniaxial spherical model.
}
\end{figure}

Our next task is to calculate explicitly the real-space 
correlation functions (\ref{cfdef}) in the small-anisotropy 
case $1-\eta \ll 1$.
Since the CFs themselves are proportional to $1/D$ and thus vanish in the 
spherical limit [see, e.g., (\ref{chitr})], it is more convenient to 
deal with the appropriate susceptibilities, which are given by
%
%\marginpar{cfrdef}
%
\begin{equation}\label{cfrdef}
\tilde\chi_{\alpha\alpha}({\bf r}) =
v_0\!\!\!\int\!\!\!\frac{ d{\bf q} }{ (2\pi)^3 }\,
e^{ i{\bf kr} } \tilde\chi_{\alpha\alpha}({\bf k}) .
\end{equation}
In the transverse case using (\ref{chitr}) one comes to the well-known result
%
%\marginpar{chitrr}
%
\begin{equation}\label{chitrr}
\tilde\chi_\perp({\bf r}) \cong 
\frac{ c_0\alpha^{1/2} }{ r } e^{ -r/\xi_{c\perp} }, \qquad
\xi_{c\perp} \cong \sqrt{ \frac{ \alpha }{ 1-\eta G} } 
\end{equation}
for $r \gg a_0$.
One can see that in the isotropic case, $\eta=1$, 
the transverse correlation length, $\xi_{c\perp}$, 
if infinite in the whole region below $\theta_c$, where $G=1$.
In more complicated situations, as for the longitudinal susceptibility 
below $\theta_c$, (\ref{chizk}), the correlation length can be determined as 
$\xi_{cz} = 1/\kappa_c$, where $\kappa_c$ is the singularity point of 
$\tilde\chi_z({\bf k})$ on the imaginary axis $k = i\kappa$, which is 
closest to the origin.
This leads with the use of (\ref{rksm}) to the transcedental equation for 
$u_c \equiv \xi_{c\perp}/(2\xi_{cz})$ having the form
%
%\marginpar{uceq}
%
\begin{equation}\label{uceq}
u_c \ln \frac{ 1 + u_c }{ 1 - u_c } = a ,
\qquad 
a \equiv \frac{ \xi_{c\perp} }{ \xi_m } 
= \frac{ 2m^2 }{ \theta c_0 \sqrt{ 1 - \eta } } ,
\end{equation}
where $\xi_m$ is given by (\ref{xim}).
This equation coincides with Eq. (4.19) of \cite{gar96jpa}, which determines
the temperature-dependent width $\delta_L=2\xi_{cz}$ of the (linear) domain 
wall in the uniaxial spherical model.
The asymptotic solutions of (\ref{uceq}) read
%
%\marginpar{ucasy}
%
\begin{equation}\label{ucasy}
u_c \cong \left\{
\begin{array}{ll}
\sqrt{a/2} ,           & a \ll 1 \;\;({\rm slightly\; below}\; \theta_c) \\
1 - 2 e^{-a} ,         & a \gg 1 \;\;({\rm far\; below} \;\theta_c) .  
\end{array} 
\right. 
\end{equation}
Note that the transition between two regimes in (\ref{ucasy}) occurs at
$\epsilon\sim\epsilon^* \sim \sqrt{1-\eta}$ [see (\ref{eps*})].
The explicit results for the correlation length $\xi_{cz}$ itself, 
including that above $\theta_c$, read
%
%\marginpar{xiczasy}
%
\begin{equation}\label{xiczasy}
\xi_{cz} \cong \left\{
\renewcommand{\arraystretch}{2.5}
\begin{array}{ll}
\displaystyle
\sqrt{ \frac{ \alpha  }{ 1-G } } \cong 
\sqrt{ \frac{ \alpha I(\eta) }{ -\epsilon } } , 
& 0 < -\epsilon \ll \epsilon^* \\
\displaystyle
\sqrt{ \frac{ \xi_m\xi_{c\perp} }{ 2 } } \cong 
\sqrt{ \frac{ \alpha I(\eta) }{ 2\epsilon } } , 
& 0 < \epsilon \ll \epsilon^* \\
\displaystyle
\frac{ \xi_{c\perp} }{ 2 } 
\left[ 1 + \exp\left(-\frac{ \xi_{c\perp} }{ \xi_m }\right)\right],
& \epsilon^* \ll \epsilon ,  
\end{array} 
\right. 
\end{equation}
where $I(\eta)$ is given by the second line of (\ref{ieta}).
The numerically calculated temperature dependences of the longitudinal and 
transverse correlation lengths are illustrated in Fig. \ref{scfxi}.
%\par
%

To calculate the Fourier-transform 
$\tilde\chi_z({\bf r})$ below $\theta_c$ with the help of (\ref{cfrdef}), 
(\ref{chizk}), 
and (\ref{rksm}), it is convenient to perform at first the integration over 
the angle variables and then to deform the $k$-integration contour 
in the complex plane.
In this way one comes to the well-behaved expression
%
%\marginpar{chizr}
%
\begin{eqnarray}\label{chizr}
&&
\tilde\chi_z({\bf r}) \cong
\frac{ 2a c_0\alpha^{1/2} }{ r }
\Bigg[
e^{-r/\xi_{cz} } \frac{ 1-u_c^2 }{ a (1-u_c^2) + 2 u_c^2 } 
%\right.
\nonumber\\
&&\qquad
%\left.
{} + \int\limits_1^\infty du \frac{ e^{-2ru/\xi_{c\perp} } }
{ \left( u \ln \frac{ u + 1 }{ u - 1 } + a \right)^2 + \pi^2u^2 }
\Bigg] ,
\end{eqnarray}
where the first term is the contribution of the pole determining the 
longitudinal correlation length $\xi_{cz}$ [see (\ref{uceq})] and the second 
one is the integral along the cut of the $\arctan$-function of (\ref{rksm}).
The expression above simplifies in various limiting cases.
For $a\ll 1$ ($\epsilon \ll \epsilon^*$), the cut term of (\ref{chizr}) is 
small as $a$ in comparison to the pole term and, additionally, it decays as 
$\exp(-2r/\xi_{c\perp})$, i.e., much faster than the first 
term of (\ref{chizr}).
Thus, in this case only the pole term in (\ref{chizr}) is relevant, and with 
the use of (\ref{ucasy}) one obtains
%
%\marginpar{chizr1}
%
\begin{equation}\label{chizr1}
\tilde\chi_z({\bf r}) \cong
\frac{ c_0\alpha^{1/2} }{ r } e^{ -r/\xi_{cz} },
\end{equation}
where $\xi_{cz}$ is given by the the middle line of (\ref{xiczasy}).
In the opposite case, $a\gg 1$ ($\epsilon \gg \epsilon^*$), 
the pole sticks to the beginning of the cut, $u=1$, and the 
amplitude of the pole term becomes exponentially small because of the factor
$1-u_c^2$ [see (\ref{ucasy})].
Thus, in this case only the cut term in (\ref{chizr}) is essential, and the 
result can be simplified to
%
%\marginpar{chizr2}
%
\begin{equation}\label{chizr2}
\tilde\chi_z({\bf r}) \cong
\frac{ 2c_0\alpha^{1/2} }{ \pi r } e^{ -r/\xi_{cz} } 
\int\limits_0^\infty \frac{ dx }{ 1+x^2 } 
\exp\left( \frac{ 2r }{ \pi \xi_m } x \right) ,
\end{equation}
where $\xi_{cz}$ is given by the the lower line of (\ref{xiczasy}).
The asymptotic forms of (\ref{chizr2}) are (\ref{chizr1}) for $r\ll 
\xi_m$ and
%
%\marginpar{chizr3}
%
\begin{equation}\label{chizr3}
\tilde\chi_z({\bf r}) \cong
\frac{ 2m^2 }{ \theta } \left( \frac{ \xi_m }{ r } \right)^2 e^{ -r/\xi_{cz} }
\end{equation}
for $r\gg \xi_m$.
One can say that for the isotropic model, where the 
true correlation length $\xi_{cz}$ is infinite below
$\theta_c$, the ``bare'' one, $\xi_m$ of (\ref{xim}), still plays its role 
in some sence: 
For $r\gsim \xi_m$ the slow decay
$\tilde\chi_z({\bf r})\propto 1/r$ changes to a faster one, 
$\tilde\chi_z({\bf r})\propto 1/r^2$.

As a conclusion, the anisotropic spherical model (ASM) considered here is a 
good exactly solvable ``toy'' model for classical spin systems, 
which can be successfully 
applied in many situations where the standard spherical model of Berlin and 
Kac fails.
More realistic models of phase transitions posess, naturally, 
the other, nonspherical, values of the critical 
indices, but this is, however, only a quantitative effect, which is 
less important in comparison to the profound role played by the 
gapless spin waves in the isotropic model below $T_c$.
The latter are properly taken into account in the ASM, which is thereby 
a very important step beyond the mean field approximation.

The author thanks D. G\"orlitz, J. K\"otzler, and Hartwig Schmidt for 
valuable discussions.
The financial support of Deutsche Forschungsgemeinschaft 
under contract Schm 398/5-1 is greatfully acknowledged.

%\bibliography{gar}

\begin{thebibliography}{10}

\bibitem{berkac52}
{T. N. Berlin and M. Kac}, Phys. Rev. {\bf 86},  821  (1952).

\bibitem{sta68prl}
{H. E. Stanley}, Phys. Rev. Lett. {\bf 20},  589  (1968).

\bibitem{sta68pr}
{H. E. Stanley}, Phys. Rev. {\bf 176},  718  (1968).

\bibitem{oka70}
{H. Okamoto}, Phys. Lett. A {\bf 32},  315  (1970).

\bibitem{barfis73}
{M. N. Barber and M. E. Fisher}, Ann. Phys. {\bf 77},  1  (1973).

\bibitem{cosmazmih76}
{G. Costache, D. Mazilu, and D. Mihalache}, J. Phys. C {\bf 9},  L501  (1976).

\bibitem{gar96jpal}
{D. A. Garanin}, J. Phys. A {\bf 29},  L257  (1996).

\bibitem{gar96jpa}
{D. A. Garanin}, J. Phys. A {\bf 29},  2349  (1996).

\bibitem{morkaw62}
{H. Mori and K. Kawasaki}, Prog. Theor. Phys. {\bf 27},  529  (1962).

\bibitem{patpok73}
{A. Z. Patashinskii and V. L. Pokrovskii}, 
Zh. Eksp. Teor. Fiz. {\bf 64} 1445 (1973)
[Sov. Phys. JETP {\bf 37},  117  (1973)].

\bibitem{lovtro91}
{S. Lovesey and K. Trohidou}, J. Phys.: Condensed Matter {\bf 3},  1827
  (1991).

\bibitem{garlut84d}
{D. A. Garanin and V. S. Lutovinov}, Solid State Commun. {\bf 50},  219
  (1984).

\bibitem{gar94jsp}
{D. A. Garanin}, J. Stat. Phys. {\bf 74},  275  (1994).

\bibitem{gar96prb}
{D. A. Garanin}, Phys. Rev. B {\bf 53},  11593  (1996).

\bibitem{holpri40}
{T. Holstein and H. Primakoff}, Phys. Rev. {\bf 58},  1098  (1940).

\bibitem{fispri85}
{M. E. Fisher and V. Privman}, Phys. Rev. B {\bf 32},  447  (1985).

\bibitem{koegoedompie94}
{J. K\"otzler, D. G\"orlitz, R. Dombrowski, and M. Pieper}, Z. Phys. B {\bf
  94},  9  (1994).

\bibitem{flogoekoe96}
{A. Flosdorff, D. G\"orlitz, and J. K\"otzler}, J. Appl. Phys. {\bf 79},  4641
  (1996).

\bibitem{gar96jsp}
{D. A. Garanin}, J. Stat. Phys. {\bf 83},  907  (1996).

\bibitem{joy72pt}
{G. S. Joyce},  in {\em Phase Transitions and Critical Phenomena}, edited by 
C.
  Domb and M.~S. Green (Academic Press, New York, 1972), Vol.~2.

\bibitem{ma76}
{S. Ma}, {\em Modern theory of critical phenomena} (Benjamin, Reading, 1976).

\end{thebibliography}

\end{document}